\documentclass[runningheads]{llncs}

\usepackage[T1]{fontenc}

\usepackage{graphicx}

\usepackage{bm}

\usepackage[hmargin=3cm,vmargin=3cm]{geometry}

\newcommand{\elabel}[1]{\label{eq:#1}}
\newcommand{\Eref}[1]{Eq.~(\ref{eq:#1})}

\newcommand{\flabel}[1]{\label{fig:#1}}
\newcommand{\Fref}[1]{Fig.~\ref{fig:#1}}

\newcommand{\Slabel}[1]{\label{sec:#1}}
\newcommand{\Sref}[1]{Sec.~\ref{sec:#1}}

\begin{document}

\title{Analysis of public transport (in)accessibility and land-use pattern in different areas in Singapore}

\titlerunning{Public transport accessibility and land use in Singapore}

\author{Hoai Nguyen Huynh\orcidID{0000-0002-8432-7435}}

\authorrunning{H. N. Huynh}

\institute{Insitute of High Performance Computing\\
Agency for Science, Technology and Research, Singapore\\
\email{huynhhn@ihpc.a-star.edu.sg}}

\maketitle
\begin{abstract}

As more and more people continue to live in highly urbanised areas across the globe, reliable accessibility to amenities and services plays a vital role in sustainable development. One of the challenges in addressing this issue is the consistent and equal provision of public services, including transport for residents across the urban system. In this study, using a novel computational method combining geometrical analysis and information-theoretic measures, we analyse the accessibility to public transport in terms of the spatial coverage of the transport nodes (stops) and the quality of service at these nodes across different areas. Furthermore, using a network clustering procedure, we also characterise the land-use pattern of those areas and relate that to their public transport accessibility. Using Singapore as a case study, we find that the commercial areas in the CBD area expectedly have excellent accessibility and the residential areas also have good to very good accessibility. However, not every residential area is equally accessible. While the spatial coverage of stops in these areas is very good, the quality of service indicates substantial variation among different regions, with high contrast between the central and eastern areas compared to the others in the west and north of the city-state. We believe this kind of analysis could yield a good understanding of the current level of public transport services across the urban system, and their disparity will provide valuable and actionable insights into the future development plans.

\end{abstract}

\section{Introduction}

With countries having pledged to reach net-zero emissions in the next few decades \cite{COP26}, plans are being put together by governments around the world to achieve the goal. Among the courses of action, a feasible way to accomplish this is to reduce the use of private vehicles and shift toward the more sustainable use of public transit. As part of the process, improving the quality and accessibility of existing public transport infrastructure is vital in achieving high ridership. Furthermore, from a social perspective, equal public transport service provision also contributes to sustainable development in urban areas \cite{2021@Ribeiro.etal,2017@Scheurer.etal,2021@Shi}.

Public transport accessibility has received much attention from various communities in the literature. Among the measures of accessibility, the public transport accessibility level (PTAL) \cite{2016@Shah.Adhvaryu}, which combines the walking distance to the transport nodes and the frequency of the services, has been frequently used for its simple calculation. However, the method's main disadvantage lies in its use of a heuristic distance threshold from a point of interest to the nearby transport nodes, which may produce artefacts in certain areas. More sophisticated methods have been proposed, such as analysing the walking distance and time to the nearest bus stop using the actual walking paths \cite{2019@Kaszczyszyn.Sypion-Dutkowska}. Yet, the approach may not be suitable for large-scale analysis where the amount of spatial data required would make the procedure computationally inefficient. Other authors have also looked at the optimal spacing of bus stops \cite{2021@Sahu.etal} so that the system's overall performance could be improved, and the link with land use could be established \cite{2016@Chen.etal,2013@Lee.etal}. Beyond the city scale, some authors have performed a comparative analysis of the share of the population within proximity to public transport in different metropolitan areas \cite{2016@Bok.Kwon}.

While public transport accessibility is intuitively about the ability and ease of users to reach their destination, it could also be examined from the opposite perspective of inaccessibility, which could be imagined to hold the hidden information about the system. Along that line, in this study, we formulate the study of public transport accessibility as an inverse problem, that is, to view the spatial inaccessible area as ``conjugate'' of its accessible counterpart. This approach has received relatively little attention in the literature, but it can be shown to provide interesting insights into the spatial organisation of an urban system. Using this approach, we will explore the public transport accessibility of different areas in terms of the bus stops' spatial (non-)coverage and their associated quality of service. The methodology will be demonstrated in the city-state of Singapore, and the accessibility of its different areas will be analysed.

In the Singapore context, several studies have been performed to analyse the accessibility to the Mass Rapid Transit (MRT) stations \cite{2019@Li.etal,2005@Olszewski.Wibowo,2005@Wibowo.Olszewski,2004@Zhu.Liu} or the impact of the expansion of the MRT network on accessibility to employment \cite{2018@Conway.etal}. Yet, few have been done for the bus network. Therefore, this work also contributes to discerning the accessibility of bus services in Singapore. While a public transport system typically contains train and bus services, it could be argued that analysing the public bus network alone is sufficient in terms of the spatial accessibility of different local areas. This is because bus stops are also well presented at the train stations, and the bus network has a high degree of penetration into the residential areas.

In the remainder of this paper, we first describe the computational method used to analyse the land-use pattern of different areas in Singapore and their associated public transport accessibility. After that, we present the results on the spatial patterns of accessibility across various regions in the country. We also offer some insights into the spatial organisation of the urban system in Singapore and discuss how the findings contribute to our knowledge of sustainable urban development.

\section{Data and methods}

\subsection{Data}

The datasets required for the analysis in this study were obtained from relevant authorities in Singapore. They can be categorised into three groups of public transport, land use and boundary of the planning areas.

\subsubsection{Public transport.}

\Slabel{bus_description}

In Singapore, public transport is managed by the Land Transport Authority (LTA), which provides a wealth of transport-related data through the DataMall portal \cite{LTA_DataMall}. For the analysis in this study, we obtained the data on the public bus services, including the information on the bus stops and the services serving those bus stops. Within the scope of this study, we only select the bus stops located within the relevant areas in Singapore (see \Sref{landuse_classification} and \Fref{landuse_classification}) for subsequent analysis. At every bus stop, we filter the data to select only the regular services. A service is considered regular if it is available throughout the day and has the lowest dispatch frequency not longer than 30 minutes.
As of January 2022, there are $5,069$ bus stops in Singapore, $4,952$ of which are served by at least a regular service. There are $553$ unique services serving all stops. Within the areas selected for analysis in this study, there are $343$ unique regular services serving $4,636$ bus stops.

\subsubsection{Land use.}

\Slabel{landuse_description}

The data on detailed land use at every land parcel in Singapore is obtained from the Urban Redevelopment Authority (URA) through the Singapore's open data portal \cite{Data_Gov}. Under the 2019 Master Plan, there are $33$ land-use types, ranging from commercial, residential to industrial use. Of these land-use types, we exclude the areas being used for ports or airports as they are generally not accessible to the public. The total area for all ports and airports excluded from the analysis in this study is about $5.5\%$ of the total area of Singapore. The list of the remaining $32$ land-use types in Singapore can be found in the label description of the plot in \Fref{landuse_characterisation}.

\subsubsection{Planning areas.}

\Slabel{planning_area_description}

The planning area data is also available from the Singapore's open data portal \cite{Data_Gov}. Under the 2019 Master Plan, there are $55$ planning areas in Singapore, each of which has a comparable population size and contain essential amenities and services required for the functions of the area and its socio-economic activities. The planning areas will serve as the spatial unit of analysis in this study in terms of both public transport accessibility and land-use patterns. $15$ of the $55$ planning areas are excluded, and the remaining $40$ areas are retained based on the land-use pattern analysis (see \Sref{landuse_classification} and \Fref{landuse_classification}). Together with the exclusion of ports and airports described in land-use data above, the area selected for analysis in this study is about $55\%$ of the total area of Singapore.

\subsection{Analysis of land-use pattern}

\Slabel{land_use_analysis}

\subsubsection{Presentation of a land-use configuration.}

Each planning area can be characterised by the types of land use that make up its area. Apart from counting how many land-use types are present within a planning area, this characterisation can also look further into the land-use composition, i.e. how much of each type is present. Because all planning areas possess different land area measures, the proportion of each land-use type is a more appropriate measure for comparison than the absolute amount of area of each land-use type. In this manner, the land-use profile of a planning area can be presented as a $32$-dimensional vector of the percentage $p_i$ of each land-use type within the planning area, with $i=1,2,\dots,32$ for the $32$ land-use types listed in \Fref{landuse_characterisation}, i.e. $\bm p=(p_1,p_2,\dots,p_{32})$. Graphically, this can also be presented as a pie chart, as shown in \Fref{landuse_characterisation}.

\begin{figure}[ht!]
\centering
\includegraphics[width=\textwidth]{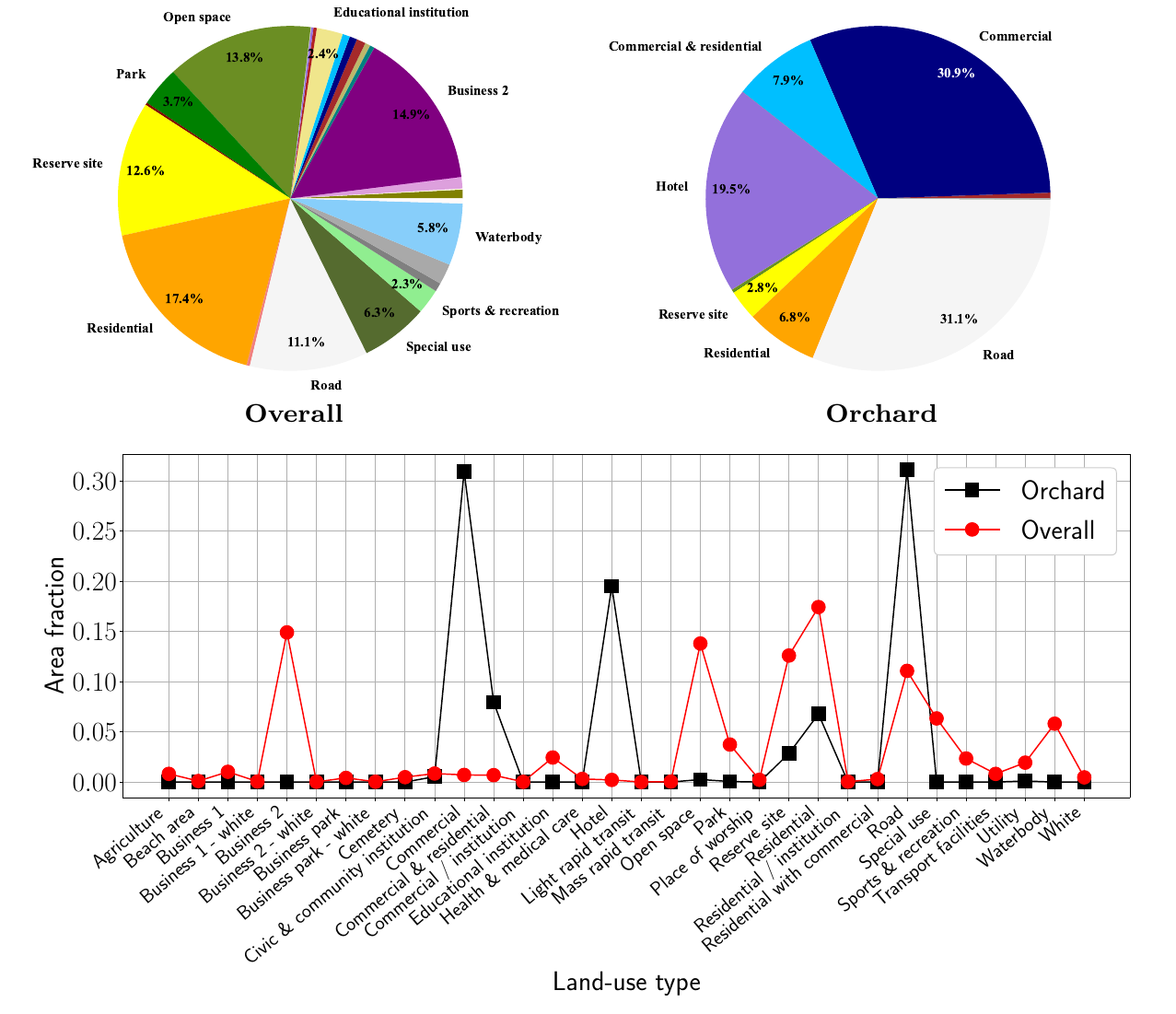}
\caption{\flabel{landuse_characterisation}Composition of land use in Singapore. \textit{Top left:} Pie chart for the entire Singapore. \textit{Top right:} Pie chart for the Orchard planning area. \textit{Bottom:} Comparison between fraction of area of different land-use types within Orchard and the entire Singapore.}
\end{figure}

\subsubsection{Similarity score between two land-use configurations.}

Geometrically speaking, the land-use configuration vector representing each planning area can be imagined to reside in an abstract $32$-dimensional space. This allows us to quantify the similarity of a pair of vectors by the (abstract) angle between them in the $32$-dimensional space. In this manner, two similar vectors would be pointing almost along a common direction, forming an angle of a small value. For quantification, cosine similarity is employed to measure how close the pair of vectors are to one another or how similar the two planning areas are. Mathematically, the following formula gives the similarity score between two vectors $\bm a$ and $\bm b$
\begin{equation}
\elabel{cosine_similarity}
\textrm{sim}(\bm a,\bm b) = \cos{\theta_{a,b}}=\frac{\bm a\cdot\bm b}{\left|\bm a\right|\left|\bm b\right|}=\frac{\sqrt{\sum_{i=1}^n{a_ib_i}}}{\sqrt{\sum_{i=1}^n{a_i^2}}{\sqrt{\sum_{i=1}^n{b_i^2}}}}
\end{equation}
in which $a_i$ and $b_i$ refer to the respective components of the vectors, represented by the percentage of the land-use types within the corresponding planning area. The similarity score ranges between $0$ (the two land-use configurations are totally different) and $1$ (they are perfectly identical).

\subsubsection{Land-use pattern clustering.}

After measuring the similarity between the land-use configuration of the planning areas, a network of planning areas can be constructed by treating each area as a node and adding a link between a pair of planning areas if their similarity score is above $0.7$. This threshold is chosen because it corresponds to an angle of 45 degrees, which geometrically suggests some degree of similarity. After this procedure, we obtain a network of planning areas in which similar ones are connected (see \Fref{landuse_community}). It can be argued that the planning area network would possess some clustering structure because some planning area nodes would have denser and stronger connections with one another, forming distinctive groups. This structure of the planning area network will be analysed using the clustering procedure as described in \cite{2022@Jiang.Huynh} to identify the groups of planning areas that are most similar to one another and classify them. The results of land-use classification will form the basis for selecting planning areas for the subsequent analysis of public transport accessibility (see \Sref{landuse_classification}).

\begin{figure}[t]
\centering
\includegraphics[width=\textwidth]{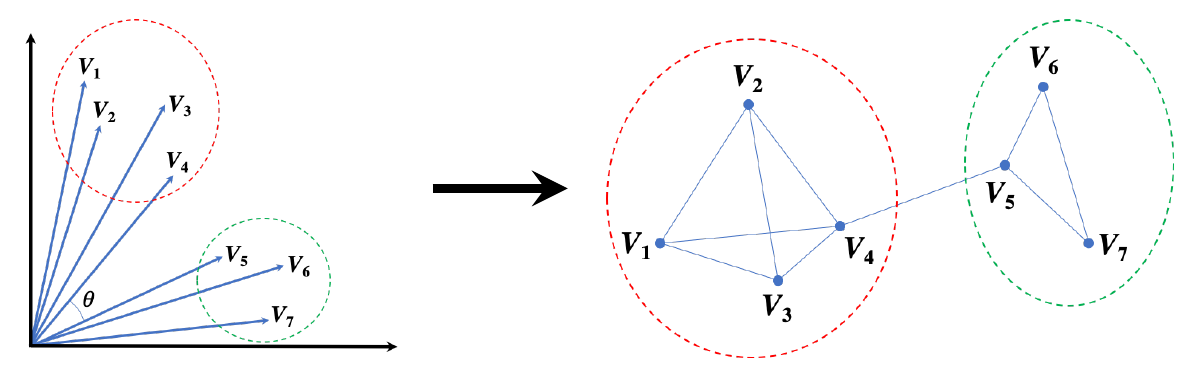}
\caption{\flabel{landuse_community}Construction of the network of land use configurations. \textit{Left panel:} land use configuration vectors. \textit{Right panel:} land use configuration network. The dotted envelops mark the identified clusters of nodes whose structure possesses the highest modularity score.}
\end{figure}

\subsection{Analysis of public transport nodes}

\Slabel{public_transport_analysis}

\subsubsection{Identification of ``non-covered'' areas.}

A place of interest is said to have good access to public transport if its distance to the nearest public transport node (bus stop) is reasonably short. Conversely, a place of interest with the nearest bus stop farther than that reasonably short distance is considered to have poor access to public transport. Using this picture, we can identify the so-called ``non-covered'' area at a given distance $\rho$, or the area that is outside (all) the circles of radius $\rho$ centred at the bus stops. For a suitable value of $\rho$, if the area covered by the buffer circles is considered the accessible area, we can call the area that is not covered by any of the circles the inaccessible one.

\begin{figure}[t]
\centering
\includegraphics[width=\textwidth]{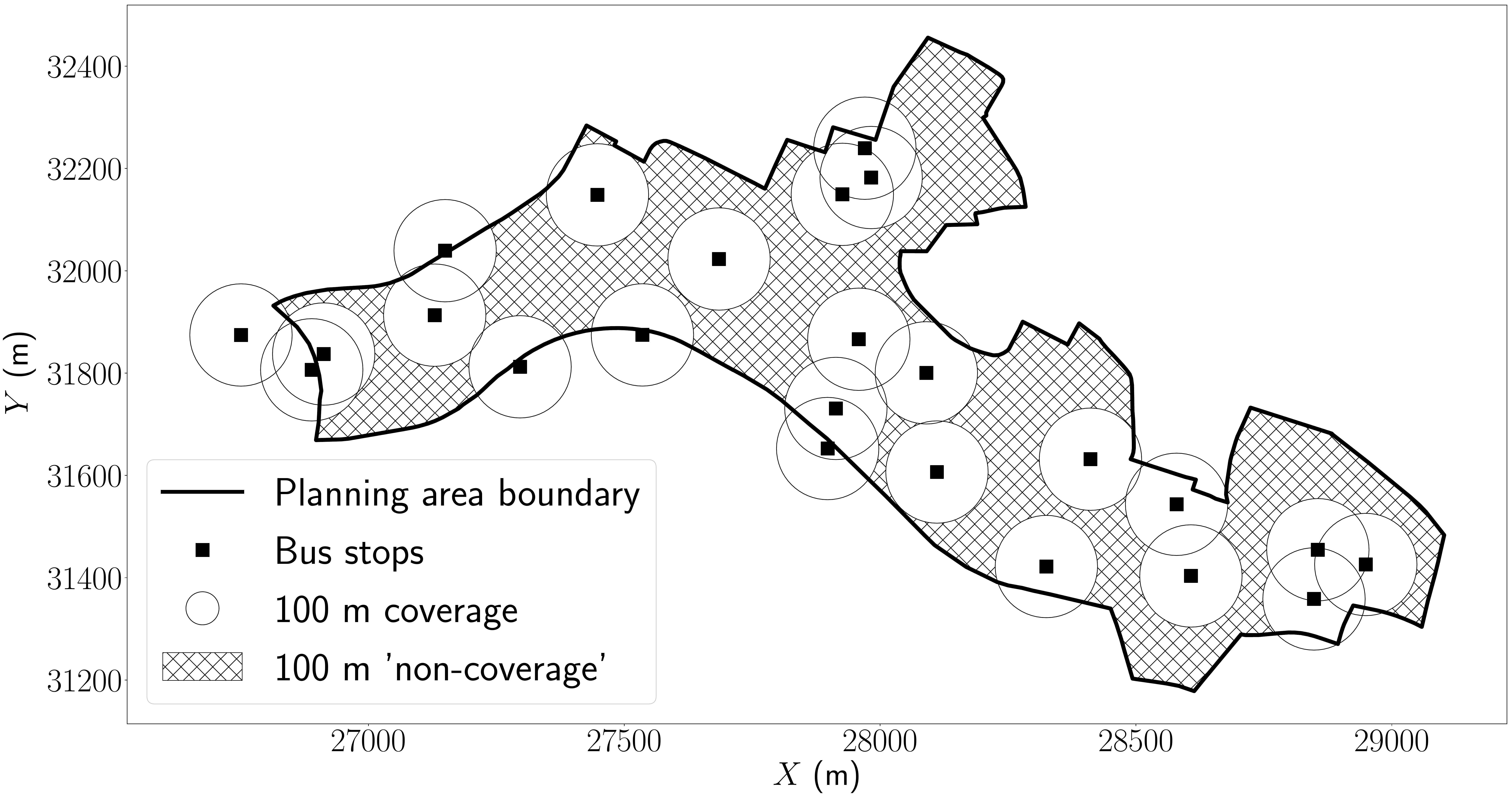}
\caption{\flabel{noncoverage_illustration}Illustration of area not covered within a buffer radius of $100$ m of the bus stops.}
\end{figure}

An illustration of identifying the non-covered area within the planning area of Orchard in central Singapore is provided in \Fref{noncoverage_illustration}. Given a distance of $100$ m, circles of radius $100$ m are drawn with centres at all the bus stops within and just outside Orchard. The union area of all these circles are created, and the difference between the area of Orchard and this union yields the non-covered area, i.e. the area of Orchard that is not within $100$ m of any bus stop.

\subsubsection{Critical distance of spatial coverage.}

\begin{figure}[ht!]
\centering
\includegraphics[width=\textwidth]{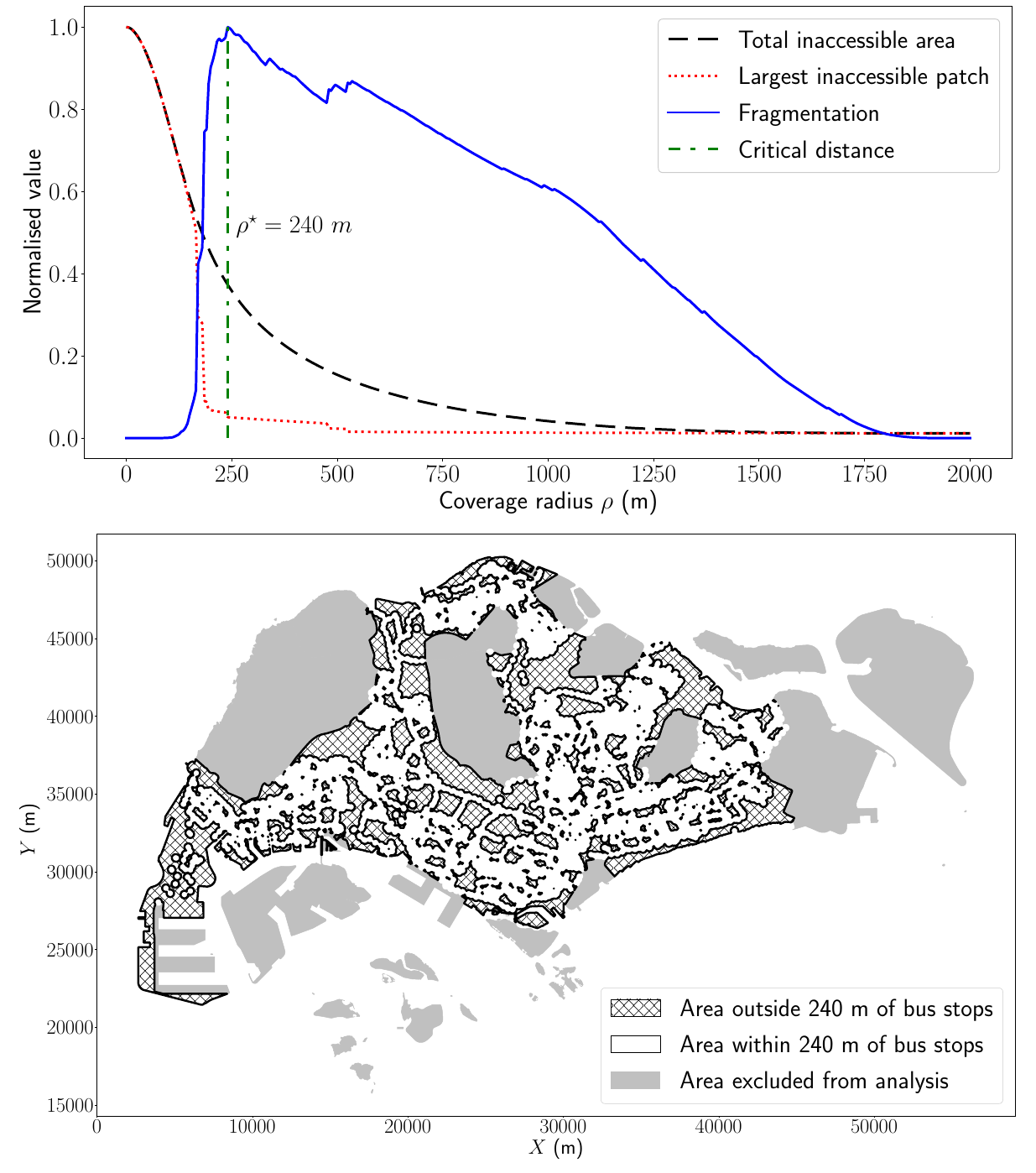}
\caption{\flabel{noncoverage_fragmentation}Fragmentation of the space not covered by the bus stops.}
\end{figure}

After determining the areas not covered within a certain radius of the bus stops, the next issue is to determine the value of the buffer radius that could be used as a suitable value to reflect the nature of accessibility to the bus stops. To tackle this question, we can look at the spatial structure of the patches formed in the process of identifying the non-covered area.

At a very small buffer distance $\rho$, pretty much all of the area of interest is not covered by the bus stops, and it exists as a connected land mass. When $\rho$ increases, the covered area starts to expand and eat away the connected land of the non-covered area. At a very large buffer distance, the reverse scenario takes place when pretty much all the area of interest is now covered by the bus stops, leaving little space not covered. In between, we can observe that the non-covered area, which started as a connected land mass, becomes highly fragmented at some intermediate value of $\rho$ before being completely destroyed, giving way to the covered areas. The notion of fragmentation comes when many patches of non-covered area are formed. In order to quantify this process, we employ the information-theoretic measure of entropy to quantify the degree of fragmentation of the non-covered patches. This is similar to the idea of measuring the complexity of a spatial point pattern \cite{2019b@Huynh} that has been applied to study the relative location of public transport nodes in the urban context \cite{2019a@Huynh,2018@Huynh.etal}.

In \Fref{noncoverage_fragmentation}, we can observe the growth curve of the non-covered area and its entropy measure. As described above, the entropy measure peaks at some intermediate value of $\rho$, which we call the critical distance. The value of the critical distance is where the inaccessible space is most fragmented, and it marks the onset of accessibility. The pattern of the non-covered area at the critical distance is also shown, exhibiting an irregular pattern that possibly resembles fractals known to exist at the critical point of transition in many physical systems \cite{2019a@Huynh}. Subsequently, we can use this critical distance to assess the spatial coverage of public transport nodes across different planning areas in Singapore.

\section{Results and discussion}

\subsection{Classification of areas based on land-use pattern}
\Slabel{landuse_classification}

While different urban areas can contain complex land-use patterns, they can generally be clustered into a small number of groups using the clustering procedure described in \Sref{land_use_analysis}. The $55$ planning areas in Singapore can be grouped into six major categories, namely commercial, residential, residential mixed with industrial, industrial, reserved, and others, based on their composition of the $32$ land-use types described in \Sref{landuse_description}. In the network of planning areas (see \Sref{land_use_analysis}), upon filtering out the links between areas whose similarity score calculated by \Eref{cosine_similarity} is below $0.7$, $9$ areas are disconnected from the main network. Inspecting the land-use profile of these $9$ areas, we find them very distinct from the rest of the areas. In particular, the Central Water Catchment (labelled \texttt{9} on the map in \Fref{landuse_classification}) and Western Water Catchment (\texttt{53}) or Lim Chu Kang (\texttt{20}) areas consist of mostly open space and waterbody. On the other hand, areas like Changi (after excluding the Changi airport) (\texttt{10}), Changi Bay (\texttt{11}), Paya Lebar (\texttt{32}) and Mandai (\texttt{21}) contain mostly land for special purposes. Conversely, the other areas of Marina South (\texttt{23}) and Southern Islands (including the popular tourist destination of Sentosa island) (\texttt{44}) are of recreational nature, with a large proportion of their land being used for parks or sports and recreation. These areas are the first ones to be excluded from the clustering procedure (as well as from the subsequent public transport accessibility analysis for their distinctive land-use characteristics) and are labelled as ``other'' land-use pattern (coloured grey in \Fref{landuse_classification}).

\begin{figure}[t]
\centering
\includegraphics[width=\textwidth]{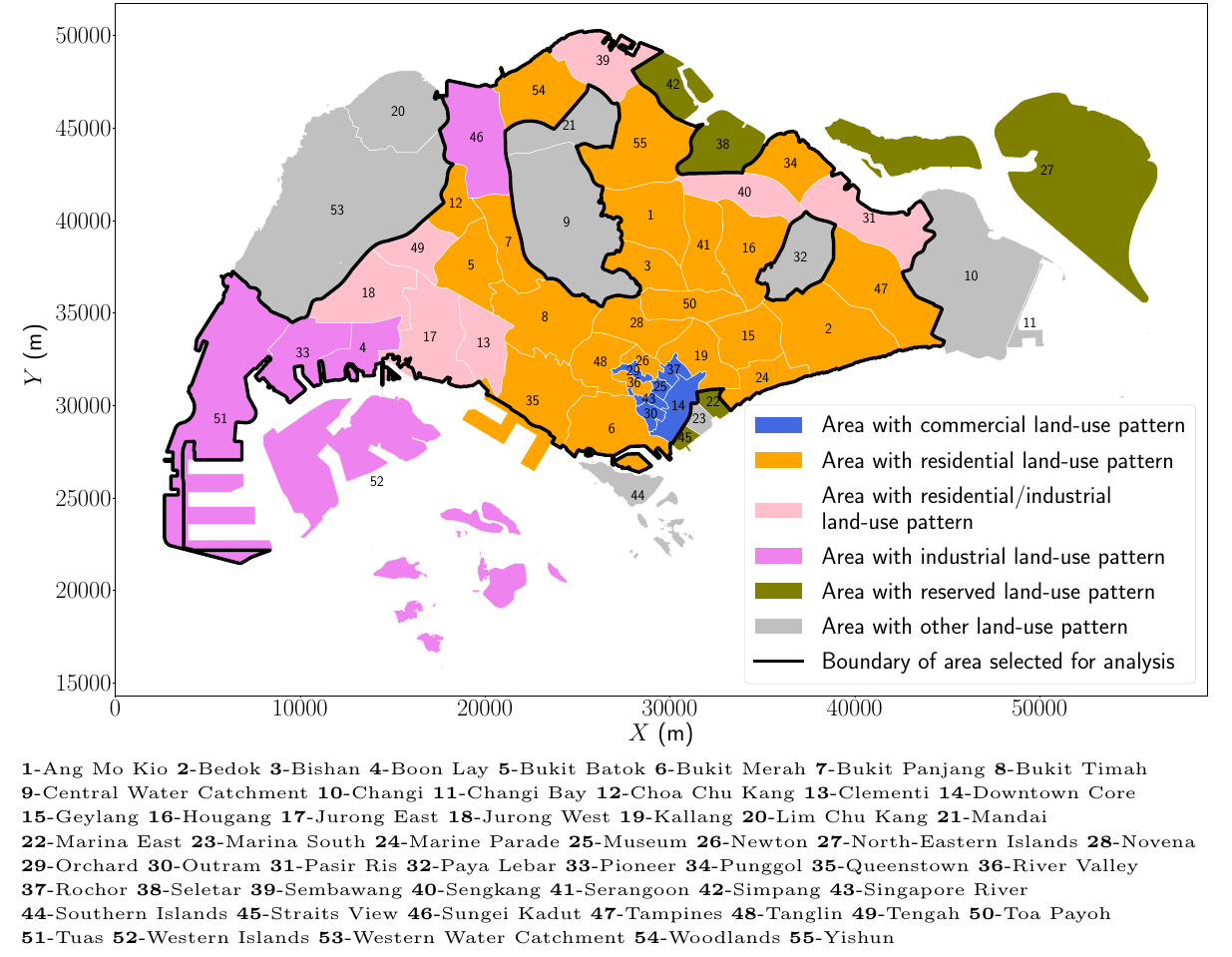}
\caption{\flabel{landuse_classification}Classification of planning areas based on composition of land-use types.}
\end{figure}

After excluding the planning areas with distinct land-use characteristics from the main network, the remaining $46$ areas are found by the clustering procedure to belong to 5 different groups. The first group contains $6$ planning areas (\texttt{14}-Downtown Core, \texttt{25}-Museum, \texttt{29}-Orchard, \texttt{30}-Outram, \texttt{37}-Rochor, and \texttt{43}-Singapore River) with a very high proportion of land for commercial and hotel usage. These areas are indeed in Singapore's Central Business District (CBD), and we label them as ``commercial'' land-use pattern (coloured blue in \Fref{landuse_classification}). The second group contains $23$ planning areas with predominantly land use of residential type and, hence, is labelled as ``residential'' land-use pattern (coloured orange in \Fref{landuse_classification}). The third group includes $7$ planning areas, which have a high proportion of land for both residential use and business 2 (clean, light industry and general industry or warehouse). These areas suggest the co-location of residential development alongside the industrial facilities and are labelled as ``residential/industrial'' land-use pattern (coloured pink in \Fref{landuse_classification}). The fourth group contains $5$ planning areas with land chiefly for business 2 and, hence, is labelled as ``industrial'' land-use pattern (coloured violet in \Fref{landuse_classification}). The final group has $5$ planning areas with land use mainly of reserved type and, hence, is labelled as ``reserved'' land-use pattern (coloured olive in \Fref{landuse_classification}).

Following the classification of the planning areas based on their land-use pattern, we decided to exclude the $5$ areas with reserved land-use pattern as these areas are reserved by the government for future development and currently do not accommodate any urban activities. Furthermore, we also exclude the area of Western Islands (\texttt{52}) as they are industrial areas and not served by the public transport system in Singapore. Together with the earlier exclusion of $9$ areas with ``other'' land-use pattern, we omit in total $15$ planning areas and retain only the remaining $40$ areas for the analysis of public transport accessibility. It should further be noted that we also exclude the ports and airports (see \Sref{landuse_description}), resulting in some planning areas like Queenstown (\texttt{35}) being cropped. The final selection of areas for analysis is shown with a thick black boundary in \Fref{landuse_classification}.

\subsection{Measure of accessibility to public transport nodes}

The analysis of the fragmentation of the non-covered patches in \Sref{public_transport_analysis} implies that the inaccessible space is most fragmented when the entropy measure maximises before being further destroyed at larger values of the buffer radius $\rho$ (see \Fref{noncoverage_fragmentation}). The value of the buffer radius when the entropy measure maximises is called the critical distance, which suggests that it is an intrinsic distance embedded within the system. Below this value, the system is largely disconnected, and above this value, the system appears to enter a different phase.

In Singapore, this critical value is found to be $240$ m and agrees reasonably well with the perception of the typical distance that most people would find comfortable for walking (3 to 4 minutes). This might reflect the result of the transport planning by the relevant authorities so that in most areas, people would walk no more than $240$ m to reach the nearest bus stops. While this value reflects an overall good spatial coverage of bus stops across Singapore, we can also utilise it to analyse and assess the public transport accessibility of different areas in the city-state.

\subsubsection{Spatial coverage of public transport nodes.}

Using the buffer distance of $240$ m, we identify the area covered by the bus stops served by at least a regular bus service (as defined in \Sref{bus_description}) in Singapore, or the ``accessible area''. We compute the total (union) accessible area within each planning area as a fraction of the total size of that planning area (excluding the area of port or airport, if any). The result of such measure over all selected planning areas is shown as a choropleth map in \Fref{accessibility_coverage}.

\begin{figure}[ht!]
\centering
\includegraphics[width=\textwidth]{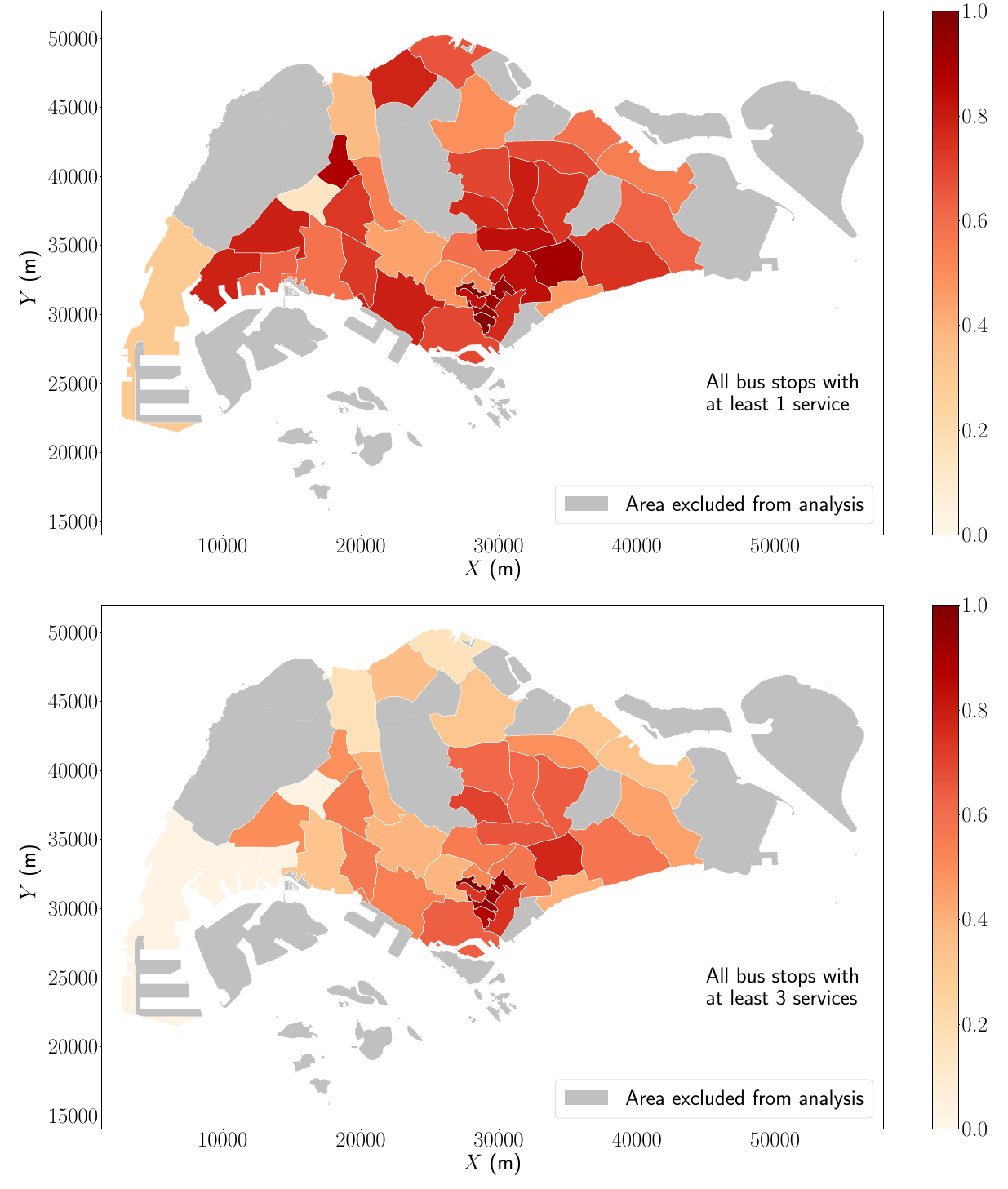}
\caption{\flabel{accessibility_coverage}Spatial coverage of bus stops at the critical distance of $240$ m in different areas across Singapore.}
\end{figure}

It could be observed that most of the planning areas selected for analysis have very good spatial coverage of the public transport nodes. Notably, the CBD has excellent coverage when the planning areas it contains are almost fully covered by the bus stops at $240$ m, indicating a high density of bus stops that people can easily access from anywhere within the CBD. Outside the city centre, the majority of the planning areas of the residential type also have a high quality of bus stop coverage, except Bukit Timah (look for label \texttt{8} in \Fref{landuse_classification} for its location in Singapore), Marine Parade (\texttt{24}), Tanglin (\texttt{48}), Yishun (\texttt{55}), and to some extent Bukit Panjang (\texttt{7}), Newton (\texttt{26}), Novena (\texttt{28}), Punggol (\texttt{34}), Tampines (\texttt{47}). The planning areas with a mixture of residential and industrial land use generally have fair coverage, with the exception of Jurong West (\texttt{18}) having very good quality and Tengah (\texttt{49}) conversely having very low quality. The remaining industrial-type areas have low spatial coverage by the bus stops except for Pioneer (\texttt{33}), having very good quality.

\subsubsection{Quality of service.}

As argued earlier, spatial coverage is only part of the measure of the accessibility of public transport in an area. Besides the spatial coverage, the quality of service at the bus stops also contributes significantly to accessibility. Bus stops served by more services are arguably more ``accessible'' as they provide travel options to more places. To probe the service quality, we look at the spatial coverage of bus stops served by at least 3 regular services, using the same buffer distance of $240$ m.

In the bottom choropleth map in \Fref{accessibility_coverage}, we can observe that the CBD still enjoys an excellent coverage of stops with at least 3 services. However, moving away from those areas, the quality of the coverage starts to decline. Yet, the decline takes different rates in different regions, with areas like Toa Payoh, Bishan, Hougang, or Geylang still having a good quality of coverage whereas their counterparts in the other regions of the country are not as covered. Across Singapore, the central and eastern parts appear to have much better accessibility than the city-state's western and northern parts.

\subsubsection{Distance scale at planning area level.}

Apart from looking at the spatial coverage of the bus stops with different service levels, we can also characterise the spatial structure of the public transport nodes in terms of the fragmentation of the non-covered areas. This can be done by computing the critical distance of the fragmentation of the non-covered patches within each planning area.

\begin{figure}[t!]
\centering
\includegraphics[width=\textwidth]{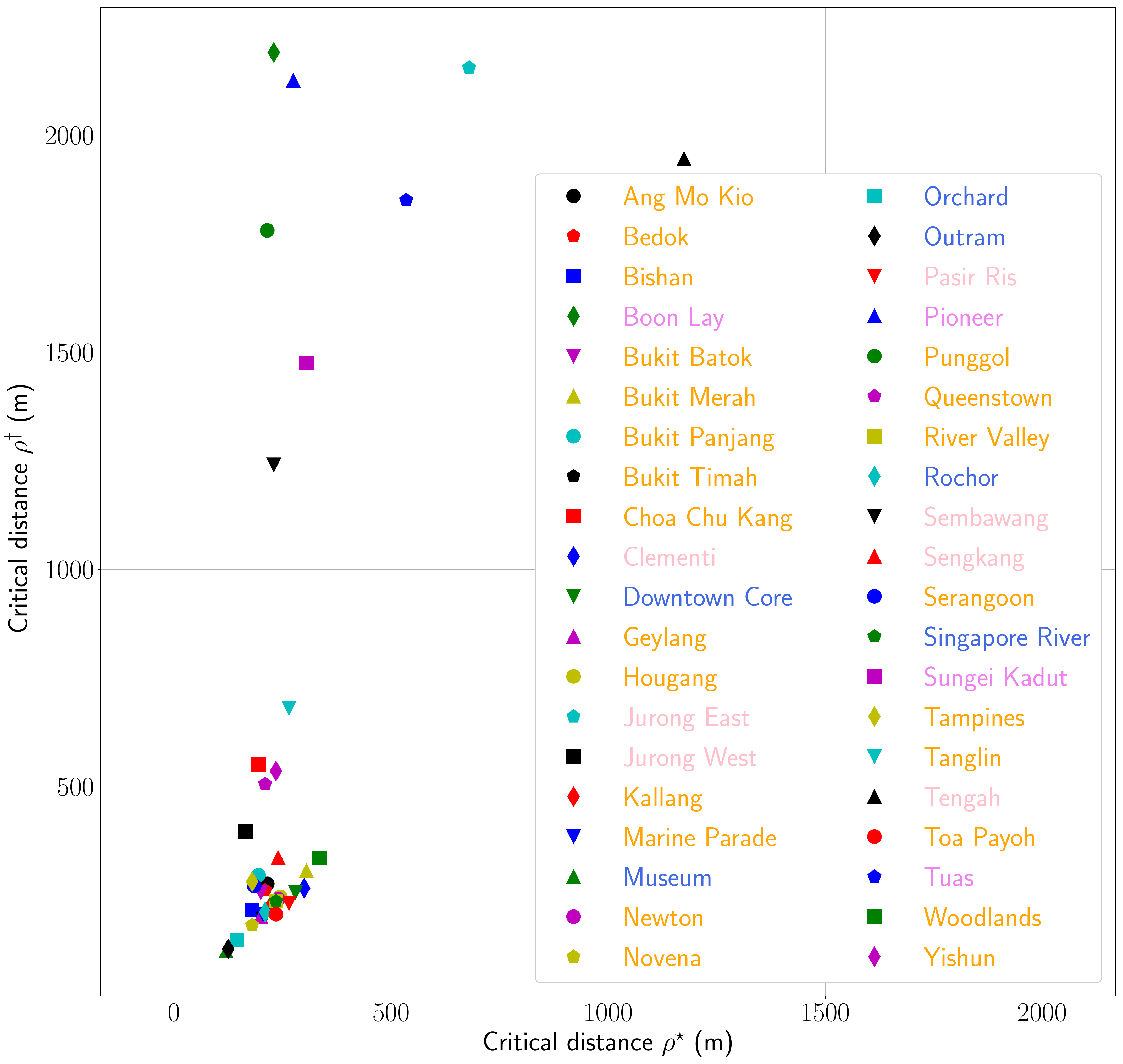}
\caption{\flabel{critical_distance_comparison}Comparison of different planning areas based on the values of their critical distance. $\rho^\star$ is for bus stops with at least 1 service, and $\rho^\dagger$ is for bus stops with at least 3 services. The name of the planning areas in the legends is coloured based on their respective land-use pattern as determined in \Fref{landuse_classification}.}
\end{figure}

\Fref{critical_distance_comparison} shows the relationship between the critical distance $\rho^\star$ computed for the fragmentation of non-covered area by bus stops with at least 1 service and the similar critical distance $\rho^\dagger$ for bus stops with at least 3 services. We can see that most areas have a small critical distance $\rho^\star$ of less than $500$ m when coverage of bus stops served by at least 1 service is considered. However, the disparity arises when we leave out the bus stops with only 2 services or less. While the majority of the planning areas still have $\rho^\dagger$ smaller than $500$ m, some of them begin to have it dilated.

\subsection{Discussion}

It should be noted that the method of geometrical analysis described in this work can also shed light on the spatial structure of an area. For example, the high fraction of area covered by the bus stops at a certain buffer radius can be related to the compactness of an area, reflecting the high density of bus stops within a small area, like the case of the planning areas in the CBD of Singapore. On the other hand, the concept of the critical distance allows us to gain insight into the sparseness of the (location of the) bus stops, which in most cases reflects the level of development \cite{2018@Huynh.etal}.

Combining the result of the spatial coverage in \Fref{accessibility_coverage} and that of the critical distance $\rho^\star$ vs. $\rho^\dagger$ in \Fref{critical_distance_comparison} could highlight the difference in the quality of accessibility across different areas in the same city, which might otherwise not be apparent. In the Singapore context, the results suggest that the new towns of Punggol and Sembawang still need much improvement for transport planning, whereas middle-aged towns like Jurong East still has a lot of potential for further development. In contrast, the areas with a high density of private housing like Bukit Timah or Marine Parade, despite having only average spatial coverage of the bus stops, appear to have a good quality of the services provided. While it is commonly understood that the residents in those areas belong to the higher income group and have a high rate of ownership of private vehicles, improving public transport accessibility in terms of shorter walking distance to bus stops could nudge their transport behaviour toward a more sustainable one.

\section{Conclusion}

In this study, we develop a computational method to analyse the quality of public transport accessibility in relation to the pattern of land use in different urban areas and apply it to the case study of the city-state of Singapore. The method combines geometrical analysis and information-theoretic measures to quantify the area that is not within a certain buffer distance of the public transport nodes, called the non-covered or inaccessible area. It is argued that the spatial structure of such inaccessible area undergoes a phase transition with the entropy measure maximising at some critical value of the buffer distance. This critical value of the buffer distance is where the inaccessible area is most fragmented, marking the onset of accessibility within the system. In Singapore, this distance is about $240$ m, indicating an overall high density of bus stops. However, analysis at the individual area level reveals that despite having good spatial coverage of the bus stops, the quality of the service at these stops varies across the country.

On the other hand, we also analyse the pattern of land use of these areas and relate it to the public transport accessibility, providing more context for the interpretation of the public transport accessibility. Typically, the commercial and residential areas of the city-state are found with very good accessibility. However, residential areas in different parts of the country exhibit marked differences, with much better accessibility in the central and eastern regions than in the west and north of Singapore. The results obtained from this research can be useful for the relevant urban and transport planning authorities in further developing the public transport network. For example, the results could help identify areas where improvements are needed and devise policies to nudge people's behaviour toward more sustainable public transport usage.

\bibliographystyle{splncs04}
\bibliography{references}

\begin{thebibliography}{10}
\providecommand{\url}[1]{\texttt{#1}}
\providecommand{\urlprefix}{URL }
\providecommand{\doi}[1]{https://doi.org/#1}

\bibitem{COP26}
{2021 United Nations Climate Change Conference}: \url{https://ukcop26.org/}

\bibitem{2016@Bok.Kwon}
Bok, J., Kwon, Y.: Comparable measures of accessibility to public transport
  using the general transit feed specification. Sustainability  \textbf{8}(3)
  (2016)

\bibitem{2016@Chen.etal}
Chen, J., Currie, G., Wang, W., Liu, Z., Li, Z.: Should optimal stop spacing
  vary by land use type?: New methodology. Transportation Research Record
  \textbf{2543}(1),  34--44 (2016)

\bibitem{2018@Conway.etal}
Conway, M.W., Byrd, A., van Eggermond, M.: Accounting for uncertainty and
  variation in accessibility metrics for public transport sketch planning.
  Journal of Transport and Land Use  \textbf{11}(1) (2018)

\bibitem{2019a@Huynh}
Huynh, H.N.: Continuum Percolation and Spatial Point Pattern in Application to
  Urban Morphology, pp. 411--429. Springer International Publishing, Cham
  (2019)

\bibitem{2019b@Huynh}
Huynh, H.N.: Spatial point pattern and urban morphology: {Perspectives} from
  entropy, complexity, and networks. Physical Review E  \textbf{100}(2),
  022320 (2019)

\bibitem{2018@Huynh.etal}
Huynh, H.N., Makarov, E., Legara, E.F., Monterola, C., Chew, L.Y.:
  Characterisation and comparison of spatial patterns in urban systems: {A}
  case study of {U.S}. cities. Journal of Computational Science  \textbf{24},
  34--43 (2018)

\bibitem{2022@Jiang.Huynh}
Jiang, Z., Huynh, H.N.: Unveiling music genre structure through common-interest
  communities. Social Network Analysis and Mining  \textbf{12}, ~35 (2022)

\bibitem{2019@Kaszczyszyn.Sypion-Dutkowska}
Kaszczyszyn, P., Sypion-Dutkowska, N.: Walking access to public transportation
  stops for city residents. a comparison of methods. Sustainability
  \textbf{11}(14) (2019)

\bibitem{LTA_DataMall}
{Land Transport DataMall}:
  \url{https://datamall.lta.gov.sg/content/datamall/en.html}

\bibitem{2013@Lee.etal}
Lee, S., Hickman, M., Tong, D.: Development of a temporal and spatial linkage
  between transit demand and land-use patterns. Journal of Transport and Land
  Use  \textbf{6}(2),  33–46 (2013)

\bibitem{2019@Li.etal}
Li, Z., Ren, S., Hu, N., Liu, Y., Qin, Z., Goh, R.S.M., Hou, L., Veeravalli,
  B.: Equality of public transit connectivity: the influence of mass rapid
  transit services on individual buildings for {Singapore}. Transportmetrica B:
  Transport Dynamics  \textbf{7}(1),  576--595 (2019)

\bibitem{2005@Olszewski.Wibowo}
Olszewski, P., Wibowo, S.S.: Using equivalent walking distance to assess
  pedestrian accessibility to transit stations in {Singapore}. Transportation
  Research Record  \textbf{1927}(1),  38--45 (2005)

\bibitem{2021@Ribeiro.etal}
Ribeiro, J., Fontes, T., Soares, C., Borges, J.L.: Accessibility as an
  indicator to estimate social exclusion in public transport. Transportation
  Research Procedia  \textbf{52},  740--747 (2021)

\bibitem{2021@Sahu.etal}
Sahu, P.K., Mehran, B., Mahapatra, S.P., Sharma, S.: Spatial data analysis
  approach for network-wide consolidation of bus stop locations. Public
  Transport  \textbf{13}(2),  375--394 (2021)

\bibitem{2017@Scheurer.etal}
Scheurer, J., Curtis, C., McLeod, S.: Spatial accessibility of public transport
  in {Australian} cities: Does it relieve or entrench social and economic
  inequality? Journal of Transport and Land Use  \textbf{10}(1) (2017)

\bibitem{2016@Shah.Adhvaryu}
Shah, J., Adhvaryu, B.: Public transport accessibility levels for {Ahmedabad,
  India}. Journal of Public Transportation  \textbf{19}(3),  19--35 (2016)

\bibitem{2021@Shi}
Shi, F.: Research on accessibility and equity of urban transport based on
  multisource big data. Journal of Advanced Transportation  \textbf{2021},
  1--18 (2021)

\bibitem{Data_Gov}
{Singapore's open data portal}: \url{https://data.gov.sg/}

\bibitem{2005@Wibowo.Olszewski}
Wibowo, S.S., Olszewski, P.: Modeling walking accessibility to public transport
  terminals: Case study of {Singapore Mass Rapid Transit}. Journal of the
  Eastern Asia Society for Transportation Studies  \textbf{6},  147--156 (2005)

\bibitem{2004@Zhu.Liu}
Zhu, X., Liu, S.: Analysis of the impact of the {MRT} system on accessibility
  in {Singapore} using an integrated {GIS} tool. Journal of Transport Geography
   \textbf{12}(2),  89--101 (2004)

\end{thebibliography}

\end{document}